\documentclass{ws-procs9x6}

\begin{document}

\title{EXOTIC $c \bar c$ MESONS}

\author{ERIC BRAATEN}

\address{Department of Physics, The Ohio State University,\\
Columbus, OH  43210, USA\\
$^*$E-mail: braaten@mps.ohio-state.edu}

\begin{abstract}
A surprising number of new $c \bar c$ mesons with masses 
above the $D \bar D$ threshold have been discovered 
at the $B$ factories.  Some of them are ordinary charmonium states, 
but others are definitely exotic mesons.  
The current theoretical status of the new $c \bar c$ mesons
is summarized.
\end{abstract}

\keywords{Charmonium; QCD; exotic mesons.}

\bodymatter

\section{New $c \bar c$ mesons}

\begin{sidewaysfigure}
\psfig{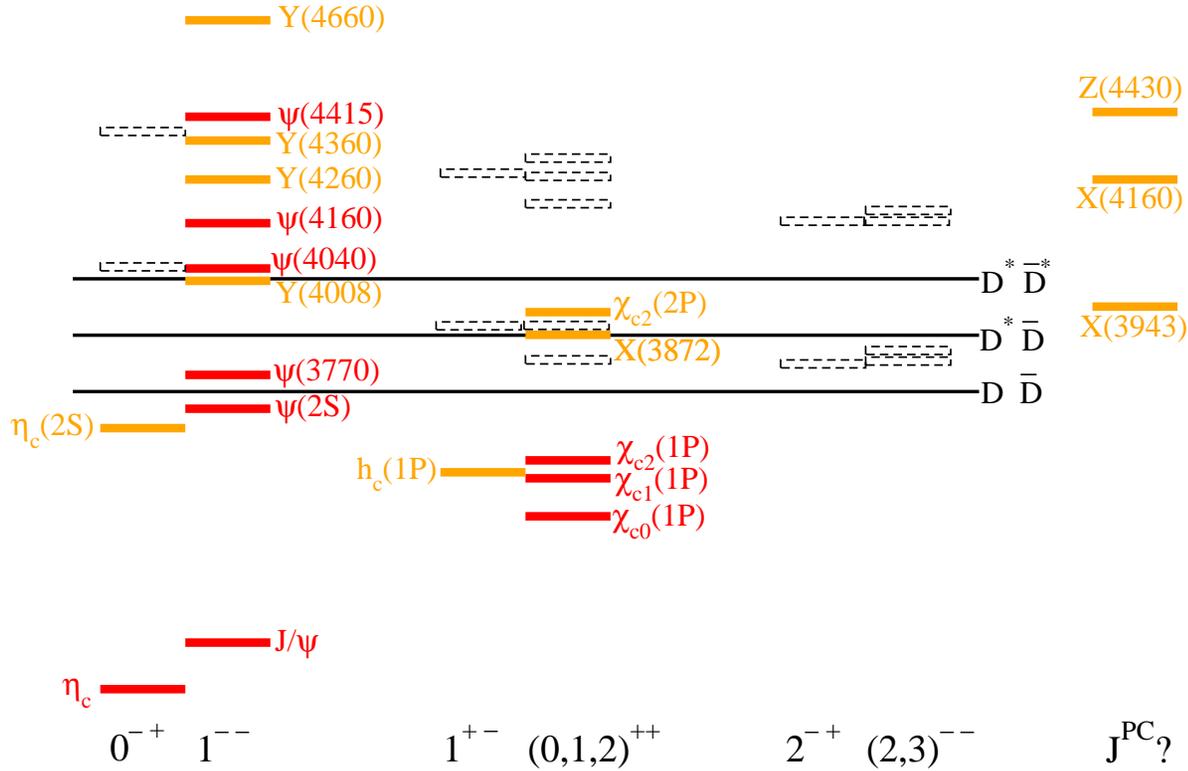}
\caption{The spectrum of $c \bar c$ mesons separated into $J^{PC}$ channels. 
The darker (red) states were discovered before the $B$ factories.  
The lighter (orange) states were discovered since 2002.
The open boxes represent missing charmonium states predicted 
by potential models.  The horizontal lines are thresholds 
for pairs of charm mesons.}
\label{fig:spectrum}
\end{sidewaysfigure}

The discovery of charmonium in November 1974 marked the beginning 
of the modern era of particle physics.
Within a few years, 10 charmonium states had been discovered.
These states are shown in Fig.~\ref{fig:spectrum}.
The charmonium states below the $D \bar D$ threshold at 3729 MeV
were the $0^{-+}$ ground state $\eta_c$, 
the $1^{--}$ states $J/\psi$ and $\psi(2S)$,
and the $0^{++}$, $1^{++}$, and $2^{++}$ states $\chi_{c0}(1P)$,
$\chi_{c1}(1P)$, and $\chi_{c2}(1P)$.
The only complete charmonium multiplet was the $1S$ multiplet 
consisting of $\eta_c$ and $J/\psi$.
The $c \bar c$ mesons above the $D \bar D$ threshold
were four $1^{--}$ states:
$\psi(3770)$, $\psi(4035)$, $\psi(4253)$, and $\psi(4421)$,
which can be identified as members of the $1D$, $3S$, $2D$, and $4S$ 
charmonium multiplets.
There were no definitive discoveries of any new $c \bar c$ mesons 
for more than 20 years.

The $B$ factories, which began operating near the beginning of the century, 
have turned out to be surprisingly effective machines for studying 
$c \bar c$ mesons.
The new $c \bar c$ mesons discovered since 2002
are shown in Fig.~\ref{fig:spectrum}.  
There are now three complete multiplets below the 
$D \bar D$ threshold: $1S$, $1P$, and $2S$.
Above the $D \bar D$ threshold 
there are four new $1^{--}$ charmonium states, 
a $1^{++}$ state called the $X(3872)$, 
the $2^{++}$ state $\chi_{c2}(2P)$,
and three states whose $J^{PC}$ quantum numbers are not yet known.
In addition, there is a charged $c \bar c$ meson called $Z^{\pm}(4430)$.
These new states provide a challenge to our understanding of QCD.
There are several recent reviews of the new $c \bar c$ mesons
\cite{Swanson:2006st,Eichten:2007qx,Voloshin:2007dx,Godfrey:2008nc}.
I will describe three of the discoveries in more detail.

\subsection{$X(3872)$}
\label{X3872}

The $X(3872)$ was discovered by the Belle Collaboration in 
November 2003 through the decay
$B^+ \to K^+ + X$ followed by $X \to J/\psi \, \pi^+ \pi^-$
\cite{Choi:2003ue}. The discovery has been confirmed by the 
CDF, BaBar, and D0 collaborations 
\cite{Acosta:2003zx,Aubert:2004ns,Abazov:2004kp}.
The mass of the $X(3872)$ is $3871.4 \pm 0.6 {\rm \ MeV}$.
Its width is less than 2.3 MeV at the 90\% confidence level
\cite{Choi:2003ue}.  Its quantum numbers are strongly preferred 
to be $1^{++}$ \cite{Abe:2005iya,Abulencia:2006ma}. 
There are two puzzling properties of the $X(3872)$ that indicate 
that it is not an ordinary charmonium state:
\begin{arabiclist}
\item
Its decays violate isospin symmetry.
The discovery decay mode $J/\psi \, \pi^+ \pi^-$ 
is dominated by $J/\psi$ and a virtual $\rho$ meson,
which is a final state with $I=1$.
However the decay of $X$ into $J/\psi \, \pi^+ \pi^- \pi^0$ 
has also been observed with a 
comparable branching fraction \cite{Abe:2005ix}.
This decay is dominated by $J/\psi$ and a virtual 
$\omega$ meson, which is a final state with $I=0$.
		 
\item
Its measured mass depends on the decay channel.
The measured mass in the $D^0 \bar D^0 \pi^0$ 
decay channel is higher 
than in the $J/\psi \, \pi^+ \pi^-$ channel by $3.8 \pm 1.1$ MeV
\cite{Gokhroo:2006bt,Aubert:2007rva}.
\end{arabiclist}

\subsection{New $1^{--}$ $c \bar c$ mesons}

Four new $1^{--}$ $c \bar c$ mesons have been discovered by the
Belle and Babar collaborations
through the initial state radiation process
$ e^+ e^- \to \gamma + e^+ e^- \to \gamma + Y$.
The $Y(4008)$ and $Y(4260)$ were discovered through their decays 
into $J/\psi \, \pi^+ \pi^-$ \cite{Aubert:2005rm,belle:2007sj}.
The $Y(4360)$ and $Y(4660)$ were discovered through their decays 
into $\psi(2S) \, \pi^+ \pi^-$ \cite{Aubert:2006ge,Belle:2007ea}.
They do not seem to be ordinary $1^{--}$ charmonium states, 
because no resonant peaks are observed in the inclusive 
$e^+ e^-$ cross section at these energies.  There are also 
no resonant peaks in the exclusive 
$e^+ e^-$ cross sections into $D \bar D$, $D^* \bar D$, 
or $D^* \bar D^*$, in spite of the masses being well 
above the thresholds for these decays.

\subsection{$Z^\pm(4430)$}

The $Z^\pm(4430)$ was discovered by the Belle Collaboration in 
August 2007 through the decay $B^+ \to K^0 +  Z^+$ 
followed by $Z^+ \to \psi(2S) \, \pi^+ $ \cite{Belle:2007wga}.
Its mass and width are $4433 \pm 5 {\rm \ MeV}$
and $45^{+35}_{-18} {\rm \ MeV}$, respectively.
Its $J^{P}$ quantum numbers are not known, 
but it has $I^G = 1^+$.
Since the decay products $\psi(2S)$ and $\pi^+$ have 
constituents $c \bar c$ and $u \bar d$, respectively,
the $Z^+$ must have constituents $c \bar c u \bar d$.
Thus $Z^\pm(4430)$ is a manifestly exotic meson.

\section{What are they?}

What are the new  $c \bar c$ mesons
above the $D \bar D$ threshold?
Some of them could be ordinary charmonium ($c \bar c$),
but some of them are definitely exotic mesons.
The exotic possibilities include
charmonium hybrids ($c \bar c g$),
tetraquark mesons ($c \bar c q \bar q)$,
and sexaquark mesons ($c \bar c q q \bar q \bar q$).
A tetraquark meson could be a compact color-singlet state,
but it could also have substructure.
The possible substructures consisting of two clusters include
charm meson molecules, diquark--antidiquark states, 
and hadro-charmonium.
I will discuss several of these possibilities in more detail.

\subsection{Charmonium}

Potential models are a phenomenological framework 
for charmonium that is well developed in most respects. 
The exception is the effect of couplings to charm meson pairs,
which are essential for accurate predictions 
above the $D \bar D$ threshold. 
Updated predictions for the charmonium spectrum from 
potential models have been given by 
Barnes, Godfrey, and Swanson \cite{Barnes:2005pb}
and by Eichten, Lane, and Quigg \cite{Eichten:2005ga}.
The missing charmonium states (and their quantum numbers) include
\begin{itemlist}
\item
one member of the $3S$ multiplet: $\eta_c(3S)$ ($0^{-+}$),
\item
three members of the $2P$ multiplet: 
$\chi_{c0}(2P)$, $\chi_{c1}(2P)$, and $h_c(2P)$
($0^{++}$, $1^{++}$, and $1^{+-}$),
\item
three members of the $1D$ multiplet:
$\psi_2(1D)$, $\psi_3(1D)$, and $\eta_{c2}(1D)$
($2^{--}$, $3^{--}$, and $2^{-+}$).  
\end{itemlist}
There are also entire missing multiplets: $4S$, $3P$, $2D$, \ldots.
The missing charmonium states are shown as open boxes 
in Fig.~\ref{fig:spectrum}.  While some of the new $c \bar c$ mesons
lie close to the predicted mass of a missing charmonium state, 
most of them do not.  

\subsection{Charm meson molecules}

Charm meson molecules are tetraquark $c \bar c$ mesons that consist 
of a pair of charm mesons ($c \bar q$ and $\bar c q$).
The interactions of the charm mesons are constrained by experimental data.
The first quantitative predictions for the spectrum of charm meson
molecules were made by Tornqvist in 1991 
\cite{Tornqvist:1991ks,Tornqvist:1994}.
He used a meson potential model with one-pion-exchange interactions 
and an ultraviolet cutoff on the $1/r^3$ potential that was tuned 
to reproduce the binding energy of the deuteron.
Tornqvist predicted charm meson molecules near threshold
in several $J^{PC}$ channels with $I=0$:
\begin{itemlist}
\item
For $D \bar D$, whose threshold is at 3729 MeV, 
there should be none.

\item
For $D^* \bar D/D \bar D^{*}$, 
there should be molecules near the threshold at 3872 MeV
in the $0^{-+}$ and $1^{++}$ channels.

\item
For $D^* \bar D^*$,  
there should be molecules near the threshold at 4014 MeV
in the $0^{++}$, $0^{- +}$, $1^{+-}$, and $2^{++}$ channels. 
\end{itemlist}
Tornqvist essentially predicted the $X(3872)$ in 1994,
although he did not anticipate the importance of the isospin splittings 
between $D^+$ and $D^0$ and between $D^{*+}$ and $D^{*0}$.

Updated predictions of a meson potential model 
have been given by Swanson \cite{Swanson:2006st}.
He tuned the ultraviolet cutoff on 
the $1/r^3$ potential to obtain the observed mass of the $X(3872)$, 
and he also included quark-exchange interactions.
Besides the $X(3872)$, whose binding energy was used as an input, 
the only other molecular state that is predicted to be bound 
is a $0^{++}$ $D^* \bar D^*$ state at 4013 MeV.
The meson potential model can also be applied to bottom meson molecules.
There should be two $B^* \bar B$/$B \bar B^{*}$ molecules
($0^{- +}$ and  $1^{++}$) and four $B^* \bar B^*$ molecules 
($0^{++}$, $0^{- +}$, $1^{+-}$, and $2^{++}$).
Swanson predicts their binding energies to range from 40 to 70 MeV.

\subsection{Tetraquark mesons}

One of the possible substructures for a tetraquark $c \bar c$ meson
is a diquark ($c q$) and an antidiquark ($\bar c \bar q$).
The problem can be dramatically simplified by treating the
constituent diquarks as point particles.
This approach has been followed by
Maiani, Piccinini, Polosa, and Riquer 
\cite{Maiani:2004vq,Maiani:2005pe,Maiani:2007vr},
by Ishida, Ishida, and Maeda \cite{Ishida:2005em},
by Ebert, Faustov, and Galkin \cite{Ebert:2005nc},
and by Karliner and Lipkin \cite{Karliner:2006hf,Karliner:2007kf}.
The most attractive color channel for a constituent  diquark
$c q$ is $3^*$.  If the diquark has no internal orbital angular momentum,
it has two possible states $S$ and $A$ corresponding to total spin 
0 and 1, respectively.
The possible S-wave tetraquarks (and their quantum numbers) are
$S \bar S$ ($0^{++}$),
$A \bar S/S \bar A$ ($ 1^{++}$, $1^{+-}$),
and $A \bar A$ ($0^{++}$, $1^{+-}$, $2^{++}$).
For each of these 6 states, there is a flavor multiplet. 
If we consider the three light quarks $q = u, d, s$, the flavor multiplet
consists of 9 states.
If we only consider two light quarks $q = u, d$, the flavor multiplet
consists of 4 states: an isospin triplet $(X^-, X^0, X^+)$ 
and an isospin singlet $X^{0'}$.

One problem with the interpretation of new $c \bar c$ mesons
as diquark--antidiquark bound states is that too many other
such states are predicted.
If we only consider diquarks with color $3^*$
and take the diquark--antidiquark system
to be in its lowest state, $6\times 9=54$ states are predicted.
There are many additional states if we also consider diquarks in the 
$6$ color state or if we allow orbital-angular-momentum
or radial excitations of the diquark--antidiquark system.  
With all these possible states, it is easy to fit the mass 
of any new $c \bar c$ meson above the $D \bar D$ threshold.  
The challenge is then to explain why 
all the other predicted states have not been observed.

One way to avoid predicting too many tetraquark states is
to start from the 4-body problem for constituent quarks.
Vijande, Valcarce, et al.  \cite{Vijande:2007fc,Vijande:2007rf}
and Hiyama, Suganama, and Kamimura \cite{Hiyama:2007ur}
have solved this 4-body problem numerically.
They have shown that there are no stable $c \bar c q \bar q$ states 
with only 2-body color-dependent forces.

A more fundamental approach to the problem 
of predicting the spectrum of tetraquark mesons is to use QCD sum rules
to determine the most attractive channels.
Navarra, Nielsen, et al.\ \cite{Matheus:2007ta,Lee:2007gs}
have shown that the QCD sum rules are consistent with 
tetraquark states in channels associated with several of the 
new $c \bar c$ mesons, including $X(3872)$ and $Z^\pm(4430)$.  

The spectrum of tetraquark $c \bar c$ mesons
should eventually be calculable using lattice QCD.
One problem is that dynamical light quarks are essential
for calculating the spectrum of $c \bar c$ mesons above 
the $D \bar D$ threshold and this makes the calculations 
computationally demanding.  The easiest masses to calculate 
are those for states with exotic quantum numbers.
The calculations for states with the same quantum numbers as 
excited charmonium states are much more difficult.

\subsection{Charmonium hybrids}

Charmonium hybrids are $c \bar c$ mesons 
in which the gluon field is in an excited state.
If the excitation of the gluon field is interpreted as
a constituent gluon, the constituents of the charmonium hybrid
are $c \bar c g$.  However there is no reason to expect the 
excitation of the gluon field to have particle-like behavior.

The spectrum of charmonium hybrids has been calculated
using lattice gauge theory.  The masses 
obtained thus far are not definitive,
because they have been calculated without dynamical light quarks.
The spectrum was calculated using a 
Born-Oppenheimer approximation by
Juge, Kuti, Morningstar \cite{Juge:1999ie}.
In this approximation, the lowest multiplet consists of 
degenerate states with the quantum numbers
$0^{-+}$, $0^{+-}$, $1^{--}$, $1^{++}$, $1^{+-}$, $1^{-+}$, 
                  $2^{-+}$, and $2^{+-}$.
Their result for the mass is approximately 4200 MeV.
The energies of charmonium hybrids with exotic quantum numbers
have been calculated using conventional lattice gauge theory by
Liao and Manke \cite{Liao:2002rj}
and by Liu and Luo \cite{Luo:2005zg}.
Their results for the masses of the lowest states 
with the exotic quantum numbers
$1^{-+}$, $0^{+-}$, and $2^{+-}$ 
are 4400, 4700, and 4900 MeV, respectively.
Since the lowest Born-Oppenheimer multiplet includes a $1^{--}$ state,
there should also be a $1^{--}$ state in this mass range.
The masses of the new $1^{--}$ states
$Y(4260)$,  $Y(4360)$, and $Y(4660)$
are compatible with this mass range.

An important selection rule for the decays of charmonium hybrids 
has been derived by
Isgur, Kokoski, and Paton \cite{Isgur:1985vy},
by Close and Page \cite{Close:1994hc}, and by
Kou and Pene \cite{Kou:2005gt}.
Decays of a charmonium hybrid into two S-wave charm mesons,
namely $D \bar D$, $D^* \bar D$, $D \bar D^*$, or $D^* \bar D^*$,
are suppressed.  Thus its dominant decay modes are expected 
to be into a P-wave charm meson and an S-wave charm meson, such as
$D_1 \bar D^{(*)}$ and $D_2 \bar D^{(*)}$.
The new $1^{--}$ $c \bar c$ mesons 
$Y(4260)$,  $Y(4360)$, and $Y(4660)$ have only been
observed in either the $J/\psi \, \pi \pi$ or $\psi(2S) \, \pi \pi$
decay channels.
Despite the large phase space available, these states 
have not been seen in the decay modes $D \bar D$, $D^* \bar D$, 
or $D^* \bar D^*$.  This makes them prime condidates for
charmonium hybrids.

\subsection{Hadro-charmonium}

Dubynskiy and Voloshin have recently proposed a new possibility 
for the substructure of tetraquark $c \bar c$ mesons:
hadro-charmonium, which consists of a charmonium $\psi$ ($c \bar c$)
and a light meson $h$ ($q \bar q$) \cite{Dubynskiy:2008mq}.
This possibility is motivated by the multipole expansion 
for the long-wavelength gluon fields that dominate the 
interaction between $\psi$ and $h$
at long distances.  They pointed out that this interaction 
is attractive and they suggested that it might be
strong enough to form a bound state.

A hadro-charmonium $\psi h$ is expected to have substantial 
decay rates into final states consisting of $\psi$ and 
decay modes of $h$.  Several of the new $c \bar c$ mesons 
have been observed to decay into final states that include
$J/\psi$ but not $\psi(2S)$ or vice versa.
This pattern is easily explained if these states are hadro-charmonia.
The $Y(4008)$ and $Y(4260)$,
which decay into $J/\psi \, \pi \pi$,
might be hadro-charmonium states containing a $J/\psi$,
while $Y(4360)$ and $Y(4660)$,
which decay via $\psi(2S) \, \pi \pi$,
might be hadro-charmonium states containing a $\psi(2S)$.
The $Z^\pm(4430)$, which decays into $\psi(2S) \, \pi^\pm$,
might be a hadro-charmonium state containing $\psi(2S)$.

\section{What is the $X(3872)$?}

The nature of the $X(3872)$ can be determined unambiguously
from the combination of two crucial experimental inputs:
\begin{arabiclist}

\item
Its mass is extremely close to the $D^{*0} \bar D^0$ threshold:
$$
M_X - (M_{D^{*0}} + M_{D^0}) = -0.4 \pm 0.7 {\rm \ MeV} .
$$

\item
Its quantum numbers are $J^{PC} = 1^{++}$.
\end{arabiclist}
The determination is based on the remarkable universal properties 
of nonrelativistic particles with short-range interactions 
that have an S-wave resonance near the threshold \cite{Braaten:2004rn}.
The universal properties are determined by the large S-wave 
scattering length or, equivalently, by the small binding energy.
The first fact above implies that the $X(3872)$ has an S-wave coupling 
to $D^{*0} \bar D^0$.  The second fact above implies that it is a 
resonant coupling.  We can conclude that $X(3872)$ is a charm meson 
molecule (bound state or virtual state) whose meson structure is
\begin{eqnarray}
X = \frac{1}{\sqrt{2}} 
\left( D^{*0} \bar D^{0} - D^0 \bar D^{*0} \right) .
\label{Xwf}
\end{eqnarray}
An example of the universal properties is a simple relation between the 
mean separation of the constituents $\langle r \rangle_X$ 
and the binding energy $E_X$: 
$\langle r \rangle_X = 1/(8 \mu E_X)^{1/2}$,
where $\mu$ is the reduced mass.
This implies that if $E_X < 0.4$ MeV, $\langle r \rangle_X > 3.5$ fm.
Thus the constituent mesons in the $X(3872)$ are usually well separated.

This determination of the nature of the $X(3872)$ provides simple 
explanations for the two puzzling features described in Section~\ref{X3872}:
\begin{arabiclist}
\item
The violation of isospin symmetry in decays of the $X(3872)$
is explained by its mass being much closer to the $D^{*0} \bar D^0$ 
threshold than to the $D^{*+} D^-$ threshold, which is higher by about 
8 MeV.  As a consequence, the meson state of the $X(3872)$,
which is given in Eq.~(\ref{Xwf}), is an equal superposition of 
$I=0$ and $I=1$.

\item
The larger mass of $X$ measured in the $D^0 \bar D^0 \pi^0$ 
decay channel compared to the $J/\psi \, \pi^+ \pi^-$ channel
is explained by the difference between the line shapes 
in those two channels.
The line shape from $B \to K + D^0 \bar D^0 \pi^0$ 
is the combination of a resonance
below the $D^{*0} \bar D^0$ threshold from $B \to K + X$
and a threshold enhancement above the $D^{*0} \bar D^0$ threshold 
from $B \to K + D^{*0} \bar D^0$ 
\cite{Braaten:2004fk,Braaten:2007dw}.
The line shape from $B \to K + J/\psi \, \pi^+ \pi^-$
has only the resonance below the $D^{*0} \bar D^0$ threshold.
Thus its peak represents the true mass of the $X(3872)$ resonance.
\end{arabiclist}
An analysis of the Belle data on $B \to K + J/\psi \, \pi^+ \pi^-$
and $B \to K + D^0 \bar D^0 \pi^0$  favors the $X(3872)$ being 
a bound state whose mass is below the $D^{*0} \bar D^0$ threshold
\cite{Braaten:2007dw}, although a virtual state whose mass is 
above the $D^{*0} \bar D^0$ threshold is not excluded
\cite{Hanhart:2007yq}.
	
Other properties of the $X(372)$ can be deduced 
by taking into account charged charm meson pairs, whose 
threshold is only about 8 MeV higher than that 
of the neutral charm meson pair
$D^{*0} \bar D^0$  \cite{Voloshin:2007hh,Braaten:2007ft}.
The ratio of the production rates of $X(372)$ in $B^0$ and $B^+$ decays
can be expressed as \cite{Braaten:2007ft}
$$
\frac{\Gamma[B^0 \to K^0 + X]}{\Gamma[B^+ \to K^+ + X]}
= \left| \frac{\gamma_1}{\gamma_1 - \kappa_1} \right|^2,
$$
where $\kappa_1 = 125$ MeV 
and $\gamma_1$ is the inverse scattering length for 
$D^* \bar D$ in the $I=1$ channel. 
This result supercedes an incorrect prediction by Braaten and Kusunoki 
that this ratio should be much less than 1 \cite{Braaten:2004ai}.
The line shape from $B^+ \to K^+ + J/\psi \, \pi^+ \pi^-$
should have a zero about 6 MeV above the $D^{*0} \bar D^0$ threshold,
while that for $B^0 \to K^0 + J/\psi \, \pi^+ \pi^-$
should have a zero about 2 MeV below the $D^{*0} \bar D^0$ threshold 
\cite{Braaten:2007ft}.  This result supercedes an incorrect prediction 
by Voloshin \cite{Voloshin:2007hh}.
In contrast, the line shapes from
$B \to K + J/\psi \, \pi^+ \pi^- \pi^0$ and
$B \to K + D^0 \bar D^0 \, \pi^0$ should have no zeroes in
the $D^* \bar D$ threshold region.

\section{Conclusions}

My main conclusions on the new $c \bar c$ mesons are as follows:
\begin{itemlist}

\item
The $X(3872)$ is a weakly-bound charm meson molecule
whose meson content is given in Eq.~(\ref{Xwf}).
It has universal properties that are determined by its 
small binding energy and are otherwise insensitive 
to details of QCD, including the 
mechanism for the binding of the charm mesons.

\item
The new $1^{--}$ mesons $Y(4260)$, $Y(4360)$, and $Y(4660)$
are good candidates for charmonium hybrids.
This interpretation could be confirmed by observing 
their decays into $D_1 \bar D^{(*)}$ and $D_2 \bar D^{(*)}$.

\item
The $Z^\pm(4430)$ is an exotic tetraquark meson
with quark content $c \bar c u \bar d$.
They are the charged members of an isospin 
multiplet $(Z^-, Z^0, Z^+)$.  The implications of this state for 
other $c \bar c$ mesons depends on
its $J^{P}$ quantum numbers, which have not yet been determined.
 
\end{itemlist}

The $X(3872)$ and $Z^\pm(4430)$ provide existence proofs
for exotic $c \bar c$ mesons.   By heavy quark symmetry,
replacing the charm quarks by bottom quarks decreases
the kinetic energy without significantly changing the 
potential energy.  This implies that exotic $b \bar b$ mesons 
should also exist and have larger binding energies.
The challenge for theory is to predict their properties
before they are discovered in experiments.

\section*{Acknowledgments}
This research was supported in part by the U.S.\ Department of Energy
under grant DE-FG02-91-ER40690.

\end{document}